# The flux dynamics behavior of the two competing high temperature superconducting phases in underdoped LaCuO$_{4.06}$

D. Di Gioacchino[a], A. Puri[b], A. Marcelli[a,c], N. Poccia[d,e], A. Ricci[f] and A. Bianconi[c,f,g,h]

[a] Istituto Nazionale di Fisica Nucleare, Laboratori Nazionali di Frascati, 00044 Frascati (RM), Italy
[b] CNR-IOM-OGG c/o ESRF LISA CRG Grenoble, 71 Avenue des Martyrs,38000 Grenoble, France
[c] RICMASS, Rome International Center for Materials Science Superstripes, via dei Sabelli 119A, 00185 Rome, Italy.
[d] NEST Istituto Nanoscienze-CNR & Scuola Normale Superiore, Pisa, Italy
[e] MESA Institute for Nanotechnology, University of Twente, P. O. Box 217, 7500AE Enschede, Netherlands
[f] Deutsches Elektronen-Synchrotron DESY, Notkestraβe 85, D-22607 Hamburg, Germany
[g] Solid State and Nanosystems Physics, National Research Nuclear University MEPhI (Moscow Engineering Physics Institute) Kashirskoye sh. 31, Moscow 115409, Russia
[h] Institute of Crystallography, CNR, via Salaria Km 29.300, Monterotondo Rome, I-00015, Italy

**Abstract**

In complex transition metal oxides (TMO) an arrested electronic phase separation (PS) appears by tuning the system near a Lifshitz transition in multiband Hubbard models. The PS in La$_2$CuO$_{4+y}$ near insulator to metal transition (IMT) is made of short range Charge Density Wave (CDW) order inhomogeneity coexisting with quenched lattice disorder. While at high doping y=0.1 percolation gives a single superconducting phase, near the IMT at y=0.06 two coexisting superconducting phases appear: the first one with a critical temperature $T_{c1}$ = 16 K and the second one with $T_{c2}$ = 29K. It is known that the two superconducting phases are characterized by two different space geometry because of two different spatial distributions of both CDW order and dopants self-organization. Here we show that these two phases show different flux dynamic regimes using alternating current (AC) multi-harmonic susceptibility experiments. This is a unique technique capable to investigate multi-phase superconductors and characterize their transport properties in a percolative scenario. Results point out that the low critical temperature phase is well described by a bulk-like flux pinning with a 2D geometry while the phase with higher critical temperature shows a 'barrier pinning' mechanism providing direct evidence of two different superconducting vortex dynamics in different complex geometrical spaces.

**Introduction**

The quenched disorder, due to spatial distribution of functional ion defects [1] and inhomogeneous misfit strain [2,3], controls complex multi-scale electronic phase separations near insulator to metal transitions (IMT) in transition metal oxides [4-8] opening the way to new emergent electronics [9]. The phase separation leads to dramatic changes of system properties, like resistivity, optical transmission and it is responsible of a variety of unusual phenomena providing unique electronic functionalities like colossal magnetoresistance materials, ferroelectrics and magnetic materials [10-12] and iron based superconductors [13].

At 30 years from the discovery of high temperature superconductivity (HTS) in transition metal oxides, the complex multiscale arrested phase separation between competing phases in a correlated quenched disorder has emerged recently as a universal feature of all families of cuprate superconductors [14-18] assigned to the intrinsic arrested phase separation near a Lifshitz transition in a multi band Hubbard system [19,20].

Cuprate superconductors are made of active superconducting layers alternated by charge storage layers that play the role of spacers, forming a hetero-structure at atomic limit with $CuO_2$ planes separated by insulating rocksalt oxide layers [21]. Many experimental observations point out that the superconductivity in cuprates emerge in a complex landscape made of puddles of nanoscale stripes [22,23] where high temperature superconductivity is made of multiple condensates in a multiband metallic phase near a Lifshitz transition [24]. The multiscale electronic phase separation in cuprates [24-27] has been observed in many HTS at mesoscopic and nanoscopic length scales [28] where it originates from a short-range inhomogeneous CDW order competing and coexisting with the high-temperature superconducting phase. In this landscape at the insulator to metal transition (IMT) the first appearing superconducting phases is the result of delocalized electronic states in percolation pathways embedded in a host space of localized electronic states [29-33].

$La_2CuO_{4+\delta}$ is the most simple cuprate perovskite showing the IMT at a critical concentration of oxygen interstitials (O-i) $\delta_c=0.055$. The mobile O-i in the crystal structure of $La_2CuO_{4+\delta}$ have a tendency toward phase separation at temperature larger than 200 K. In particular, in the insulating phase of $La_2CuO_{4+\delta}$ with a doping range of $0.01<\delta<0.055$, a phase separation occurs between two different structural phases: an oxygen-poor insulating and antiferromagnetic phase and a minority Fmmm oxygen-rich superconductive phase [34-37]. Since a similar phase separation is not observed in the cation-doped $La_{2-x}Sr_xCuO_4$, its origin has been associated to the high mobility of the oxygen interstitials (O-i). In the metallic phase at higher doping levels (i.e., for $0.055<\delta<0.12$) only one crystalline structure (Fmmm) is detected, but the superconducting transition exhibits two well-defined steps at 15 K and around 30 K pointing out the occurrence of an electronic phase separation in two distinct superconducting phases [38,39]. Using scanning micro X-ray diffraction [40] it has been shown that in the metallic phase there is a phase separation between puddles of striped ordered O-i embedded in a host space of disordered O-i [40]. More intriguingly, in the $La_2CuO_{4+\delta}$ it has been showed that the two phases of O-i self organization can be manipulated by quenching [41]. In the quenched sample the single superconducting phase can be restored by annealing in a controlled way and, this process provides an efficient tool to investigate the phase separation. The procedure consists in an annealing of the sample at 370 K, followed by a quenching below 200 K. It yields to a mixed state with two critical temperatures [40,41]: a first phase with $T_{c1}$ =16 K and a second phase with $T_{c2}$ around 32 K. Therefore, the two superconducting phases coexisting in the sample are associated with different percolation geometry in different portions of the same crystal. Indeed, annealing and quenching procedures strongly affect the formation of ordered oxygen interstitials superstructures with a highly inhomogeneous distribution at the micrometer scale. Moreover, it has been shown that it is also possible to control and manipulate locally the O-i arrangement by x-ray illumination [40]. Recent experimental data point out the presence of striped charge density wave (CDW) ordered puddles at the nanometer scale competing/coexisting with puddles of ordered O-i [42] like in Hg1201 [28]. Magnetic measurements, such as Direct Current (DC) susceptibility as a function of both temperature and DC field have been previously used to probe both the $LaCuO_4$ antiferromagnetic (AF) phase ($T_n$ ~250 K due to a spin-canted out of the $CuO_2$ plane) [43] and the two superconducting phases arising in the overdoped $LaCuO_{4+\delta}$ system via thermal quenching [33, 35, 37, 38].

While it is well known that in underdoped phase of cuprate perovskites, doped by mobile oxygen interstitials, the system is in the verge of a chemical phase separation where in some case, two different underdoped superconducting phases coexist no investigation on different pinning mechanisms in the two competing phases have been published. Since recently new electronic and structural scanning probes have shown that the inhomogeneity extends at nanoscale and mesoscale where the vortex lattice pinning takes place, the characterization of the different pinning mechanisms is now mandatory. For this purpose, we propose to use the technique of the 'magnetic multi-harmonic susceptibility" combined with the analysis of Cole-Cole plots of the third harmonic, to clearly recognize the dominant pinning type mechanism in the two phases observed in the investigated superconducting $La_2CuO_{4+y}$ single crystal. This original approach shows how the dynamics of transport-driven flux quanta is correlated-with/driven-by different nano/micro-scale texture geometries of striped domains of oxygen interstitials (O-i) and striped charge density wave (CWD) puddles present in the sample. No other techniques have been used before to correlate the vortex dynamics with the sample morphology. This original approach has a more general point of view, it can be used to monitor the evolution and the redistribution of nano- and micro-structures activated by different mechanisms such as heating, thermal quenching, doping, x-ray photo induction or other set-ups applied to functional materials that have a measurable magnetic response.

The low-frequency AC complex susceptibility dynamic technique has been already used to investigate dissipative effects in high-temperature superconductors [44]. It is well known that in presence of an external oscillating magnetic field, the coefficients $\chi_n$ of the susceptibility can be regarded as the Fourier coefficients of the steady magnetization cycle [44,45]. In particular, the real part of the first harmonic, $\chi'_1$, probes the diamagnetic shield, while the imaginary part of the first harmonic, $\chi''_1$, is a measure of all losses [46]. Actually, the high harmonic coefficients, generally described by a critical state model [47] probe only the non-linear flux-pinning response of the superconducting phases [45, 48-57]. Its generalization is based on a flux creep dynamic approach of the non-linear magnetic flux diffusion equation in superconductors [45, 55-59].

This experimental approach has the following advantages: i) the signal of the first harmonic contains information due to non-linear processes of the superconducting phases and to linear processes associated to losses that limit the superconducting state (i.e., the free motion of the quantum flux in a liquid-like configuration and of unpaired electrons). On the contrary, the high harmonics contributions to the magnetic susceptibility depend only by non-linear processes associated to the superconducting phases (e.g., correlations, pinning, etc.);

ii) if only one ac field is applied, because of the symmetry of the sinusoidal excitation field the even harmonics are not present;

iii) among the high harmonics, the third one is that with the most intense signal.

In this work, using the AC multi-harmonic magnetic susceptibility we show the occurrence of different flux pinning dynamics in different domains of the same single crystal, due to a self-organization of oxygen interstitials [60]. The sample measured is a thin $La_2CuO_4$ single crystal with a surface of 2x2 mm grown by the flux method. The oxygen interstitials were inserted in the crystal by electrochemical oxidation. The samples were fully charged using a constant potential of 0.6 V (vs. Ag(AgCl) at a charging rate of ~100 mA/cm$^2$ for some days. The weight gain has been measured by the thermo-gravimetric analysis (TGA). The tritation results show that the saturated sample doping is p= 0.15 (±0.02) with a single critical temperature of 40 K. The underdoped samples with a reduced oxygen content p=0.06 (±0.01) was obtained by heating the sample in He gas and controlled by a TGA scan or by heating the sample in vacuum above 500 K. The variation of the lattice parameters by diffraction has been used to confirm the oxygen stoichiometry 4.06 (±0.01). The orthorhombicity parameter z= 2(b-a)/(b+a) =0.45 is smaller than both that of the stoichiometric ($T_c$=0 K) sample with z=0.8 and of the fully oxygenated sample ($T_c$=40 K) with z=1.15. The result is confirmed by the c-axis lattice parameter c=13.18 Å for the y=0.06 sample, which is smaller than the fully oxygenated sample y=0.11 sample with c=13.20 Å, but larger than the stoichiometric y=0 antiferromagnetic sample c=13.14 Å [61].

The two coexisting phases in the $La_2CuO_{4.06}$ system ($T_{c1}$ = 16 K and $T_{c2}$ = 29 K) show different behaviors in the susceptibility third harmonic component as a function of temperature and amplitude of the magnetic field and underline two diverse flux dynamics, which are in turn connected with the coexistence/competition between different O-i ordered superstructures in $La_2CuO_{4+\delta}$ that is associated to different percolative mechanisms [29-32]. Actually superconducting pathways with a fractal nature form metallic percolative networks. The equilibrium among these coexisting phases can be tuned by external parameters like temperature, magnetic field, pressure, etc.

While a *bulk-like flux pinning with a 2D geometry* can be associated to the low temperature phase ($T_c$ ~ 16 K) exhibiting a stronger superconducting flux pinning response [62], the high-temperature phase ($T_c$ ~ 29 K) is better described by a *barrier-like pinning* with a thin strip geometry [63]. Data also point out a strong similarity among the AC susceptibility third harmonic pattern of the $La_2CuO_{4+\delta}$ low-$T_c$ phase and data of other Y123 superconductors [52] where 3D-dimensionally ordered CDWs emerge also at high magnetic field [64]. On the contrary, the high-$T_c$ phase behavior is similar to that observed in the Bi2223 [50].

**Results and discussion**

The sample magnetic susceptibility has been measured by a first derivative configuration gradiometer based on a bridge of two pick-up coils connected in series, wounded in the opposite sense and surrounded by a driving excitation coil [52,65]. Zero magnetic field cooling (ZFC) and magnetic field cooling (FC) temperature experiments have been performed to probe the pinning force. The trapped flux is indeed the difference between the amount of the excluded (ZFC) and expelled flux (FC). This quantity is a measure of the pinning intensity, in particular higher is the trapped flux and greater is the pinning intensity. An opposite behavior is a clear indication of a weak pinning [66]. In the upper panel of figure 1, we show the real part of the first harmonic as a function of temperature for an applied magnetic field of 5.4 G at the frequency of 107 Hz.

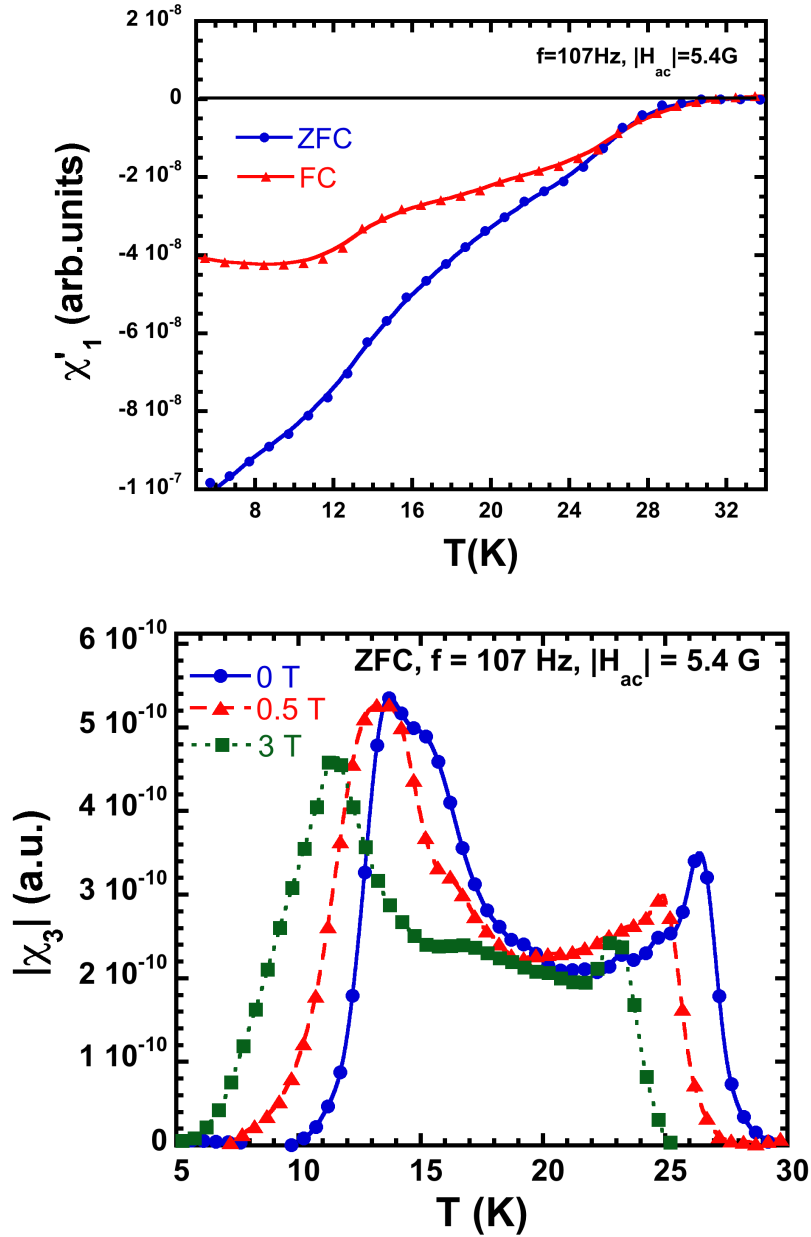

**Figure 1 Upper panel**: Comparison between the real part of the first harmonic ($\chi'_1$) of the AC magnetic susceptibility, collected in both ZFC and FC modes vs. temperature (f = 107 Hz for both procedures). **Lower Panel**. Comparison of the modulus of the third harmonic $|\chi_3|$ of the AC magnetic susceptibility vs. temperature for different amplitudes of the DC magnetic field.

In the lower panel we compare the module of the measured AC susceptibility third harmonic components $|\chi_3|$ vs. temperature, for different DC magnetic fields. Data allows to recognize two superconducting phases at $T_{c1}$ = 16 K and $T_{c2}$ = 29 K. The $T_c$ of the two phases present in this sample can be obtained from the two peaks in the derivative of the magnetic susceptibility. In this work, we have evaluated the $T_c$ superconducting transitions from the asymmetric lorentzian fit onsets calculated on the third harmonic maximum values of the superconducting phases (to clarity not shown in the lower panel of Fig. 1). Looking at the data, the comparison with the real part of the first harmonic points out a difference in the onset of the transition of 1 K. This variation is associated to the transition from the superconducting glass phase (i.e., the onset of the third harmonic) to the flux liquid phase (i.e., the onset of the first harmonic).

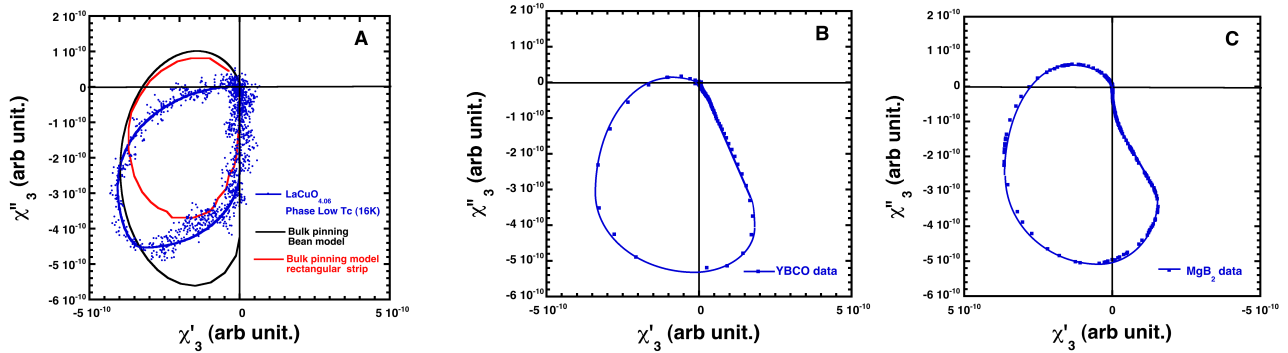

**Figure 2. (panel A)** Cole-Cole plots of the low-$T_c$ phase of the $La_2CuO_{4+\delta}$ single crystal (blue line); Bean 'Bulk pinning' theoretical model in ref. 53, fig.6 and ref.48 table I, (black curve); strongly non-linear numerical analysis of magnetic flux diffusion equation in rectangular strip with d/W =0.33 with perpendicular magnetic field (ref. 62, fig.6), d and W are thickness and width. This equation is based on a bulk pinning with a the diffusion coefficient equal $E(j)= E_c(J/J_c)^n$ with a large exponent n, n=51 (red curve). **(Panel B)** experimental Cole-Cole plot of YBCO in ref. 52, fig.7B (blue line). **(Panel C)** experimental Cole-Cole plot of $MgB_2$ in ref. 45, fig.7A (blue line). In B, C panels, the experimental data are scaled to comparison of the $La_2CuO_{4+\delta}$ data showed in panel A.

For the phase at higher temperature ($T_{c2}$ = 29 K) both FC and ZFC behaviors are similar pointing out a weak pinning mechanism. On the contrary, for the phase at 16 K the FC and the ZFC curves are well separated, a condition consistent with a stronger pinning property. In particular, the weight of the contribution of the high $T_c$ phase at temperatures below the low $T_c$ phase (~16 K) can be estimated as ~2e-8 (arb. units - see the upper panel in figure 1), while the contribution of the low $T_c$ phase measured at ~6 K, removing the high $T_c$ phase contribution, is roughly two times larger: 4e-8 (arb. units - see the upper panel in figure 1).

The third harmonic analysis allows to enhance the sensitivity to the non-linear flux pinning dynamic processes of the two superconducting phases and to better identify micro-structural pinning defects. Indeed, compared with the other harmonics, the third harmonic component of the susceptibility offers a better signal to noise ratio, discriminating directly between linear losses and non-linear flux-pinning response. This panel clearly shows two peaks corresponding to the two superconducting phases present in the $La_2CuO_{4+\delta}$. As expected, increasing the DC magnetic field both phases shift towards lower temperatures and their amplitudes decrease. However, the peak temperature shift, $\Delta T_p$, is greater for the phase at higher $T_c$ pointing out that the magnetic field has a larger effect on this phase. This is consistent with its weaker pinning respect to the low temperature phase. If we define $\delta = \Delta T/\Delta H$, we have $\delta_{p2}=1.09$ K/T and $\delta_{p1}=0.83$ K/T for the high- and low- temperature phase, respectively. The different trend of $T_p$ confirms the occurrence of different flux pinning behaviors and dynamic responses of the two superconducting phases.

To gain more information on the flux pinning dynamic it is very useful to represent the third harmonic components of the AC susceptibility in the complex-plane, i.e., $\chi''_3$ versus $\chi'_3$, also called Cole-Cole plot [67]. Actually, different flux pinning mechanisms yield to different polar plots, which may discriminate if the superconducting response has a bulk or a barrier pinning behavior. Recognizing the pinning type it is also important to give an evaluation of the sample geometry of the ratio between the width and thickness (d/W) thus to have a more effective comparison among experimental data and numerical analysis available in the literature [61,62]. In our LaCuO$_{4.06}$ sample we have W≈3d, or (d/W≈0.33).

In the figures 2 and 3 we show separately the third harmonic Cole-Cole plots for the low-$T_c$ (blue) and high-$T_c$ (blue) phase of the La$_2$CuO$_{4+\delta}$ single crystal. In the polar plane the low-$T_c$ phase placed in the third quadrant shows a "lens" shape, while the high-$T_c$ phase located in the third and fourth quadrants has an elliptical shape. Respect to the high-$T_c$ phase curve, the low-$T_c$ one encloses a larger area, confirming its stronger flux pinning characteristic.

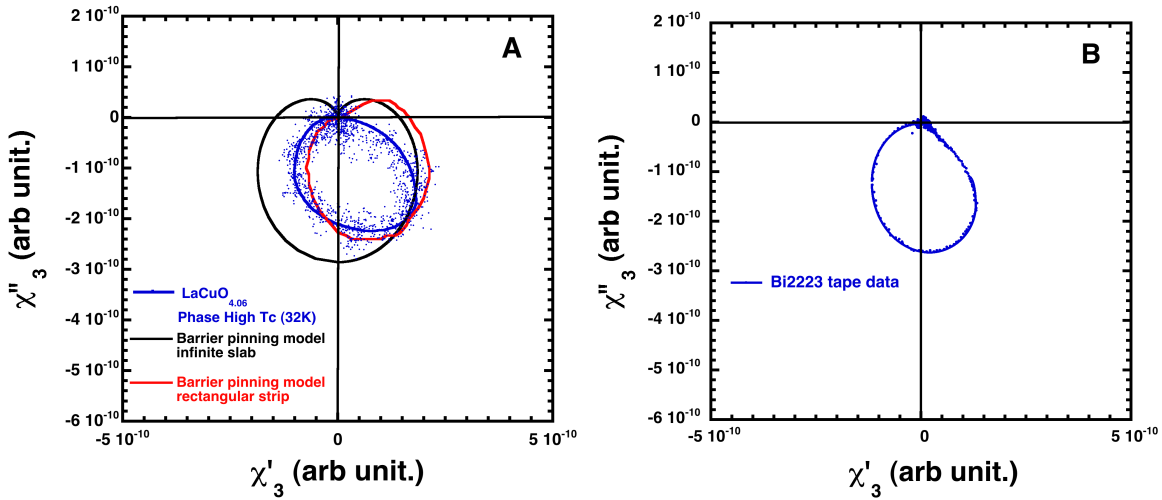

**Figure 3 Panel A:** Cole-Cole plots of the high-$T_c$ (blue) phase of the La$_2$CuO$_{4+\delta}$ single crystal. Black curve shows ideal Bean critical state due to the barrier pinning edge-shape with infinite slab in parallel magnetic field (ref.51 fig.7); red curve refers to critical state model based on a barrier pinning edge-shape configuration with rectangular d/W=0.2 (Ref.62, fig 7(b)) where d and W are thickness and width of the slab. **Panel B:** Experimental Cole-Cole plot of Bi2223 from fig.5 in ref. 50 (blue line). Experimental data are scaled to compare La$_2$CuO$_{4+\delta}$ data showed in the panel A.

To support this claim we compare in the panel A of figure 2, the Cole-Cole plot of the low-$T_c$ phase with an ideal 'bulk Bean pinning model' in a parallel field (black curve) due to the response of an infinite slab (n->∞ and W->∞)[52, 55, 62, 63], and a strongly non-linear flux creep diffusion coefficient with bulk-like pinning occurring in a rectangular strip of finite size (red curve), with ratio d/W=0.33 where d and W are thickness and width, the exponent n = 51 shows a strongly non-linear diffusion coefficient $E(j)= E_c(J/J_c)^n$ under a perpendicular field [62]. Both calculations of the response occupy the third (and partially the fourth) quadrant, in good agreement with the behavior of the low-$T_c$ phase we observed. Data clearly point out that the low-$T_c$ phase of the underdoped $La_2CuO_{4+\delta}$ single crystal is compatible with a bulk screening current and an almost bulk-like pinning [52, 55, 63, 67] within a geometrical striped configuration.

In figure 3 Panel A, we compare the Cole-Cole plot of the high-$T_c$ phase of the $La_2CuO_{4+\delta}$ single crystal (blue curve) with theoretical calculations for an ideal 'barrier pinning edge-shape' due to the response of an infinite slab in a parallel field (black curve) and a 'barrier pinning edge-shape' occurring in a rectangular strip with ratio d/W=0.2 where d and W are thickness and width under a perpendicular field (red curve) [63]. The shape of the Cole-Cole plot of the rectangular strip case is similar to the $La_2CuO_{4+\delta}$ single crystal high-$T_c$ phase. The comparison points out that the flux-pinning behavior of the $La_2CuO_{4+\delta}$ high-$T_c$ phase is similar to the flux dynamic response of a superconductor with a geometrical finite dimension of the strip, dominated by a barrier-pinning mechanism. We pointed out that the mismatch among theoretical models and experimental data of Cole-Cole plots are due to the fact that theory is based on the Bean model. The latter is based on a critical current independent by both magnetic field and time, a condition fulfilled only under particular conditions. Actually, the "real" dynamics and the time response of the quantum flux affect the shape of the Cole-Cole plots. To be more clear, if the pinning of the phase at high $T_c$ contributes to the Cole-Cole plot of the low-$T_c$ phase, we should observe a cardioid shape extending also in the 4$^{th}$ quadrants, a condition which is lacking in all panels in Fig. 2. Moreover, the fit in Figure 3 (from reference [62]) showed the Cole-Cole plot for the ratio d/W in the range 0.005 <d/W <0.2. Although the ratio does not correspond to our d/W value (0.33), the trend is in good agreement with our analysis.

In addition, because in the $La_2CuO_{4+\delta}$ the phase separation occurs within a striped domain scenario we point out that for the high-$T_c$ phase, the 'surface barrier pinning' mechanism within a thin striped geometric structure is connected to the local lattice distortion striped domains characteristic of the $La_2CuO_{4+\delta}$, in particular, to the observed CDW domains [31]. Moreover, if we consider that the low-$T_c$ phase is reasonably described by a bulk-like flux pinning within a slab geometry, as the result of the superstructure organization and ordering of the O-i, its pinning pattern is compatible with the presence of the nano-scale Q2 O-i striped grains present in the sample [40,41]. In this scenario, we had to underline that isolated pinning structures diffused in the sample may result in a pinning bulk-like contribution [68] and a percolative behavior associated to 3D puddles or bubbles of superconducting domains [42].

To additionally support this interpretation, a comparison between the third harmonic components of the two $La_2CuO_{4+\delta}$ phases and other strongly layered Bi2223 data in the panel B of Figure 3. Because the magnetic response of a superconductor is proportional to the sample volume, we discuss only the shape of Cole-Cole plots, which have been proportionally rescaled to allow the comparison. From the comparison with figure 3 panel B, it is evident that the non-linear superconducting response of the $Bi_2Sr_2Ca_2Cu_3O_{10+x}$ (Bi2223) is quite similar to the curve of the $La_2CuO_{4+\delta}$ high-$T_c$ phase (figure 3 panel A). The result points out a structural similarity between the Bi2223 superconductor and the $La_2CuO_{4+\delta}$ high-$T_c$ superconducting phase and their similar stripe morphology. On the contrary, the $YBa_2Cu_3O_7$ (YBCO, blue) sample in figure 2 panel B and the $MgB_2$ (figure 2 panel C) can be compared with the low-$T_c$ phase of the $La_2CuO_{4+\delta}$ (figure 2 panel A), in agreement with the bulk-like flux pinning observed in the YBCO, and due to a stripe-like superstructure associated to distributed grains in the sample. Moreover, compared with the $La_2CuO_{4+\delta}$ low-$T_c$ phase (figure 2 panel A) and the ideal bulk pinning (figure 2 panel B), the trend of the Cole-Cole $\chi_3$ loop of YBCO is due to a different phase lag of the magnetic signal in the sample. However, the shape of the Cole-Cole plot is rotated counterclockwise in the polar plane, a behavior compatible with a different non-linear flux creep dynamics. Although with a small different phase lag respect to YBCO, data exhibit a similar behavior with data in Fig. 6 of Ref. [53].

**Conclusions**

We have showed that the AC multi-harmonic susceptibility technique is a suitable and almost unique method to characterize the vortex pinning behavior of two coexisting phases in the underdoped $La_2CuO_{4+\delta}$. This cuprate with $T_{c1}$ = 16 K and $T_{c2}$ = 29 K is indeed characterized by different flux dynamic properties, pinning strengths and percolative mechanisms. In particular, the low-$T_c$ phase of the underdoped $La_2CuO_{4+\delta}$ has strong pinning characteristics with a *bulk-like pinning* behavior and a 3D-like percolative behavior. The high $T_c$ phase exhibits a weaker pinning strength being characterized by a *surface barrier pinning* mechanism and a bidimensional percolative behavior associated to superconducting puddles.

We sincerely acknowledge the technical staff of the DAΦNE-Light laboratory. A special thanks is due to F. Tabacchioni for his continuous technical support in the LAMPS laboratory.